\documentclass[twocolumn,showpacs,preprintnumbers,amsmath,amssymb]{revtex4}
\usepackage{graphicx}% Include figure files
\usepackage{dcolumn}% Align table columns on decimal point
\usepackage{bm}% bold math
\usepackage[usenames]{color}

\begin{document}
%Title of paper
\title{Simulation of neutrino and  charged particle production and propagation in the atmosphere}
\author{L. Derome}
%\author{M. Bu\'enerd}
\affiliation{Laboratoire de Physique Subatomique et de Cosmologie,  53, avenue des Martyrs - Grenoble, France\\}

\begin{abstract}
A precise evaluation of the secondary particle production and
propagation in the atmosphere is very important for the atmospheric
neutrino oscillation studies. The issue is addressed with the
extension of a previously developed full 3-Dimensional Monte-Carlo
simulation of particle generation and transport in the atmosphere, to
compute the flux of secondary protons, muons and neutrinos. Recent
balloon borne experiments have performed a set of accurate flux
measurements for different particle species at different altitudes in the
atmosphere, which can be used to test the calculations for the
atmospheric neutrino production, and constrain the underlying hadronic
models. The simulation results are reported and compared with the
latest flux measurements. It is shown that the level of precision
reached by these experiments could be used to constrain the nuclear
models used in the simulation. The implication of these results for
the atmospheric neutrino flux calculation are discussed.
\end{abstract}

\pacs{Valid PACS appear here}
\maketitle

\section{Introduction}

The need of a precise knowledge of the atmospheric production of secondary particles induced by the 
cosmic ray (CR) flux of charged particles, became a major stake for the scientific community with the  
recently reported evidence for atmospheric neutrino oscillations pointing to a non-zero neutrino mass, 
which evidence was based on a comparison of the measured data with secondary atmospheric neutrino flux 
calculations.
In this context, the rapidly increasing amount and statistical significance of the data collected by 
underground neutrino detectors \cite{superkandsoudan,soudan}, also made precise 
atmospheric neutrino flux calculations highly desirable as a potential tool to test calculations and models. 
One-dimensional calculations of the neutrino flux have been considered for long as a good approximation of 
the flux at earth, until some first attempts of three dimension calculations were performed 
\cite{honda,battistoni,barr}, which rapidly became the new standard approach. A survey of the different 
approaches can be found in \cite{wentz}.
A precise calculation of the neutrino flux relies on a precise knowledge of the hadronic production cross 
sections, of nucleons and mesons at the top of the decay chains leading to neutrinos in the final state. 
These cross sections however are not known in general with a high level of accuracy. 

A three-dimensional (3D) simulation describing the CR induced cascade in the atmosphere, particle propagation 
in the geomagnetic field, and interactions with the medium, was developed by the authors for the 
interpretation of the AMS01 data. The code was successfully used to reproduce the proton, electron-positron, 
and helium 3 flux data \cite{DER0,DER1,DER2} measured by AMS and their respective dependence on the geomagnetic
coordinates. The lack of a 3D calculation of the neutrino flux at this time led the authors to be the first to 
investigate this important issue either, with some first results reported in \cite{LI00}. A complete report on 
the calculated muon and neutrino flux was published in \cite{munu}. 

This paper reports the results of a further investigation of the issue, with the calculations improved by 
several respects. One improvement consisted in the use of variance reduction techniques (see section 
\ref{sec:vrt} below) which allowed, together with some code optimization, to appreciably increase the statistics 
of the simulated events sample, and thereby improve the Monte-Carlo
precision (statistical accuracy) of the calculation. A second
improvement was based on the observation that the particle production
cross sections are the main source of uncertainty in the secondary
particle flux calculations. The reliability of the neutrino flux
simulation is usually tested on the capacity of the calculations to
reproduce the atmospheric muon and proton flux measurements.  The
latter can be as well used to directly constrain the secondary
particle production cross sections, and to reduce the simulation
uncertainties on the calculated neutrino flux.  This prospect has been
explored in details and it is shown (see section~\ref{sec:atmfit}) that
the accuracy of the flux calculations can be significantly improved by
this method.

The paper is organized as follows. The method and models used in the calculations are introduced in section 
\ref{sec:simu}. Section \ref{sec:results} is devoted to the results on the proton, muon and neutrino flux in 
the atmosphere. Summary and conclusions are given in section \ref{sec:conc}.

\section{Simulation model}
\label{sec:simu}
The calculation proceeds by means of a full 3D-simulation Monte Carlo simulation. In this section the main 
features of the program are recalled for convenience. See refs \cite{DER0,DER1,DER2,BA03,munu} for other details.

\subsection{Cosmic ray flux}

For the primary flux, the 1998 AMS measurement of CR proton and helium flux \cite{PROTHELI} and the fit from 
\cite{WS} for heavier elements (up to iron), are used. The kinetic energy range of incident CRs covered in the 
simulation is $[0.2, 10000]$~GeV/n.

For each periods of the solar cycle, the incident cosmic flux are corrected for the different solar modulation 
effects using the simple force field approximation \cite{SOLMOD}.

Incident Cosmic Rays are generated on a virtual sphere at a finite distance in the earth neighborhood.
The flux at any point inside the volume of this virtual sphere is isotropic provided the 
differential element of the zenith angle distribution of the particle direction generated on the sphere is 
proportional to $\cos\theta_z\ d(\cos\theta_z)$, $\theta_z$ being the zenithal angle of the particle. 
The geomagnetic cut-off is applied by backtracing the particle trajectory in the geomagnetic field (using 
the method described in \ref{Propagation}), keeping in the sample only those particles reaching a backtracing
distance of 10 Earth radii. 
The altitude of the virtual sphere has to be chosen outside the atmosphere to ensure a backtracing process in a
region free of interaction. In the present case it was chosen at 2000~km.

For cosmic ray nuclei, the superposition model \cite{engel} was used. In this approximation a nucleus of mass 
number $A$ and charge number $Z$ is replaced by $A-Z$ neutrons and $Z$ protons with the same velocity as the 
parent nucleus. The particles are then processed like real protons and neutrons, excepted that the effective 
charge used for the propagation in the geomagnetic field is set to $Z/A$ to keep the same rigidity (i.e. the 
same trajectory) as the parent nucleus.
\begin{figure}
\begin{center}
\includegraphics*[width=0.4\textwidth]{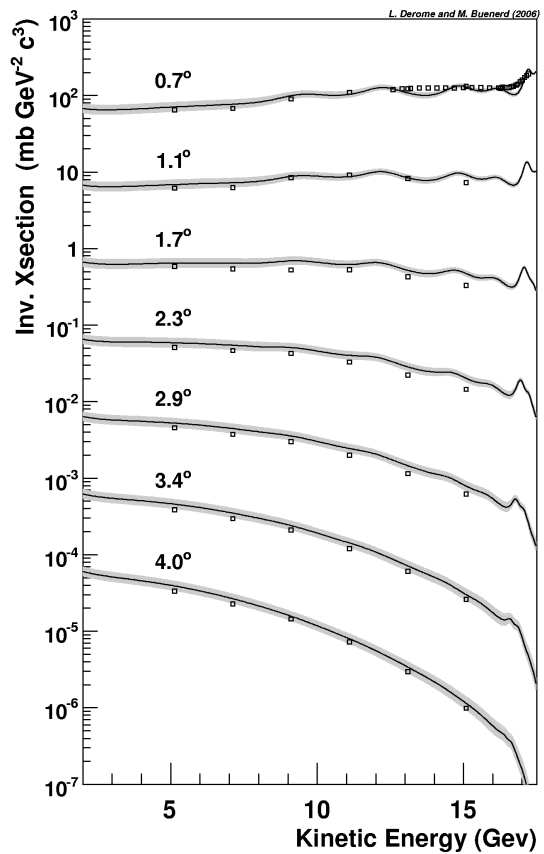}
\caption{\label{fig:Allaby} 
%Left : 
Proton invariant triple differential cross section measured (open square) 
in the $p + Be\rightarrow p +X$ at 19 Gev/c reaction \cite{allaby} compared with the fitted parametrization as
a function of the proton kinetic energy and for different production angles : 0.7$^o$, 1.1$^o$ (scaled by 
a factor $10^{-1}$), 1.7$^o$ ($\times 10^{-2}$), 2.3$^o$ ($\times 10^{-3}$), 2.9$^o$ ($\times 10^{-4}$),
3.4$^o$ ($\times 10^{-5}$) , 4.0$^o$ ($\times 10^{-6}$).
}
\end{center}
\end{figure} 
\subsection{Atmospheric and Geomagnetic model}

The model used to simulate the atmosphere is the MSISE-90 model available from 
\cite{MSISE} which describes the  temperature and densities in the earth's 
atmosphere from ground to thermospheric altitudes.

The geomagnetic model used in this version of the simulation is IGRF9 \cite{IGRF} which 
is the ninth generation of a mathematical model of the Earth's main field and its annual 
rate of change (secular variation) adopted by the International Association of Geomagnetism
and Aeronomy (IAGA).

\subsection{Particle propagation \label{Propagation}}

Each particle is propagated in the geomagnetic field and interacts with nuclei of the local 
atmospheric density. The particles trajectory is built by numerically integrating the equation 
of motion using the fifth order Runge Kutta method with adaptive step-size control from 
ref~\cite{recipes}. At each step of the propagation, in addition to the three coordinates
and the three components of the momentum, the grammage crossed by the particle is computed. 
This allows the integration step size to be adapted to the local value of the atmosphere 
density and then insure a smooth penetration in the atmosphere and a precise calculation of 
the interaction probability and of the ionization energy loss, and the latter be computed 
for each step along the trajectory.
	
Every secondary particles are processed the same way as their parent particle, leading
to the generation of an atmospheric cascade, more or less extensive, for each event. Nucleons, 
pions and kaons are produced for each interaction with their respective triple differential 
cross sections.

\subsection{Particle interaction and secondary production}
 \label{sec:param}

The cross sections used to produce secondary particles ($p$, $n$, $\pi^\pm$, $K^\pm$) are 
a very important input in the calculation, and the main source of uncertainty in the 
atmospheric flux estimation. In the present approach the particle production cross sections 
are obtained from fits to the data using an approach based on the Kalinovsky-Mokhov-Nikitin 
(KMN) parametrization of the inclusive hadronic cross sections \cite{kali}.
In the present work, a wide set of experimental data have been used to constrain the parameters
of a modified KMN analytical formula \cite{Cross}. The set of data included in the fits 
have been selected to cover as far as possible, the kinematics of interest for atmospheric 
secondary particles production.
An interest of this method is that together with the best fit parameter set for each reaction 
channel, the errors on the parameters can also be estimated.

An illustrative example is shown on figure \ref{fig:Allaby} with the
comparison between the data from \cite{allaby}
($p+\textrm{Be}\rightarrow p+X$ at 19 Gev/c) and the fit results
(solid line). For all the studied samples of each reaction channel,
the reduced $\chi^2$ obtained from fits were found larger than 1,
typically between 1 and 5. This could be expected since the
statistical and systematic errors on the cross section measurements
are usually approximate or poorly known. To account for this
uncertainty, a scale factor on the experimental error equal to
1/$\sqrt{\chi^2}$ has been included to evaluate the 95\% confidence
interval on the cross sections. This procedure is strictly valid only
in case of purely statistical errors. It is thus not completely
satisfying from this point of view but it should be rather
conservative however. The gray band on the figure corresponds to this
95\% confidence interval obtained for the proton production cross
section (from the overall proton data analysis).

The obtained confidence intervals on the cross sections were then used to compute the uncertainty 
on the secondary particles flux, and ultimately, on the neutrino flux. 

For the proton induced kaon production and the pion induced pion production the original KMN 
parametrizations were used \cite{kali}. These cross sections are less critical since their 
contribution to the final neutrino flux are rather small, namely less than 25\% and 5\% 
respectively up to 30~GeV incident energy.

\subsection{Particle decay}
In the simulation, the ($\nu, \bar{\nu}, e^{\pm}$) spectra from muon decay, were generated 
according to the Fermi theory, and the muon polarization was taken into account as in the 
previous work by the authors. For the kaon decay, the Dalitz plot distribution given in 
\cite{pdg04} was used.

\subsection{Variance reduction techniques}
\label{sec:vrt}

To increase the efficiency of the simulation several techniques could be used to reduce the 
variance of the estimator produced by the simulation \cite{vrt}. The technique used here 
was the important sampling method which consists simply of increasing arbitrarily the incident 
flux of particles, where the latter produce secondaries which contribute dominantly to the 
final differential flux to be evaluated, in a particular kinematical domain. 

For instance, in the computation of the high energy neutrino flux, the high energy galactic 
flux of charged cosmic rays provides the dominant contribution. The natural abundance of this
generic flux is extremely small however. The above mentioned technique is then used to get 
around this difficulty and the HE CR production probability is enhanced by a factor increasing the production 
yield of the secondaries of interest. The incident particle and all the particles produced in 
this event will then simply be weighted with the inverse enhancement factor used to boost the 
particle production. This weight is then taken into account when computing the estimator from 
the simulation (see next section). 

Another classical use of this method is for computing the flux of a given particle in a given 
area at a given altitude. The primary particle flux generated vertically downwards around the 
zenith of the fiducial virtual detector area will have a much larger contribution to the 
calculated final flux, compared to CRs produced far away from this geographical region. 
In this case a natural way to improve the simulation efficiency is to increase production 
probability for particles generated near the detection region, and to correct for the induced 
bias by appropriate statistical weighting of the corresponding events, as described above.

The advantage of the method is that all contributing particles are taken into account in  
the generation process, none of them being excluded or neglected. A bad choice in the 
enhancement factor used to increase a certain class of primary particles would thus increase the 
variance of the estimators but it would not bias the results.
\subsection{Particle detection and flux estimation}
Each particle was traced by the program, and his history, trajectory parameters and kinematics,
were recorded. The event file was then analyzed separately to generate the various distributions 
of interest.

To compare the simulated and the measured flux, a virtual surface of detection was defined at 
the detector altitude. It had to be large enough to ensure a large counting statistics and a 
precise estimation of the physics observables, but small enough to preserve the accuracy of the 
calculation. In practice the size of area for each computation was chosen to provide an accuracy 
below 1\%.

The simulated differential particle flux is calculated 
($m^{-2}\cdot s^{-1}\cdot sr^{-1}\cdot GeV^{-1}$) from the sum of the number of particles detected 
in the chosen area times, for each particle, the geometrical efficiency in the given particle direction 
of the considered experiment, times the weight obtained from the sampling method discussed in section 
\ref{sec:vrt} to compensate the enhancement factor.

This latter number is then divided by the (local) surface of the particle collection area, the 
integral of detector acceptance over the full solid angle, energy bin size, and equivalent sampling 
time of the CR flux.

The equivalent sampling time of CR flux is obtained from the total event number generated in the 
simulation run(s) divided by the surface of the generation sphere times the integrated 
($\cos\theta_z$ weighted) solid angle ($\pi$) times the energy integrated flux.
\begin{figure}[h]
\begin{center}
% scp ccali:~/group/laurent/cosmic/bess99/ana/Bess99Proton.eps .
\includegraphics[width=0.35\textwidth]{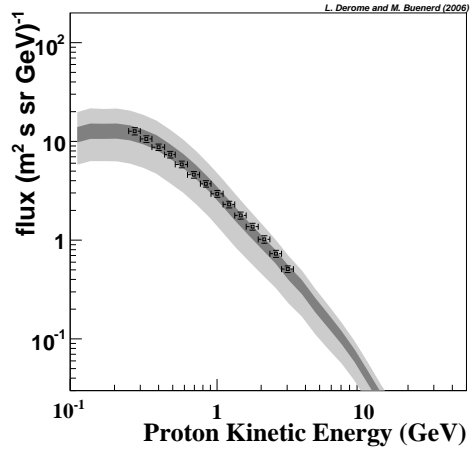}
\caption{\label{bessProton} Open squares : Proton flux measured by the BESS experiment at mountain altitude 
2770 m above sea level, data from \cite{BESS99}.  Light gray band : 95 \% confidence interval from the 
simulation. The estimation of the confidence interval includes only errors from the production cross sections. 
The dark gray band corresponds to the 95~\% confidence interval obtained 
by fitting the cross section on atmospheric data (see section \ref{sec:atmfit}.}
\end{center}
\end{figure} 
\begin{figure*}
\begin{center}
\includegraphics[width=0.75\textwidth]{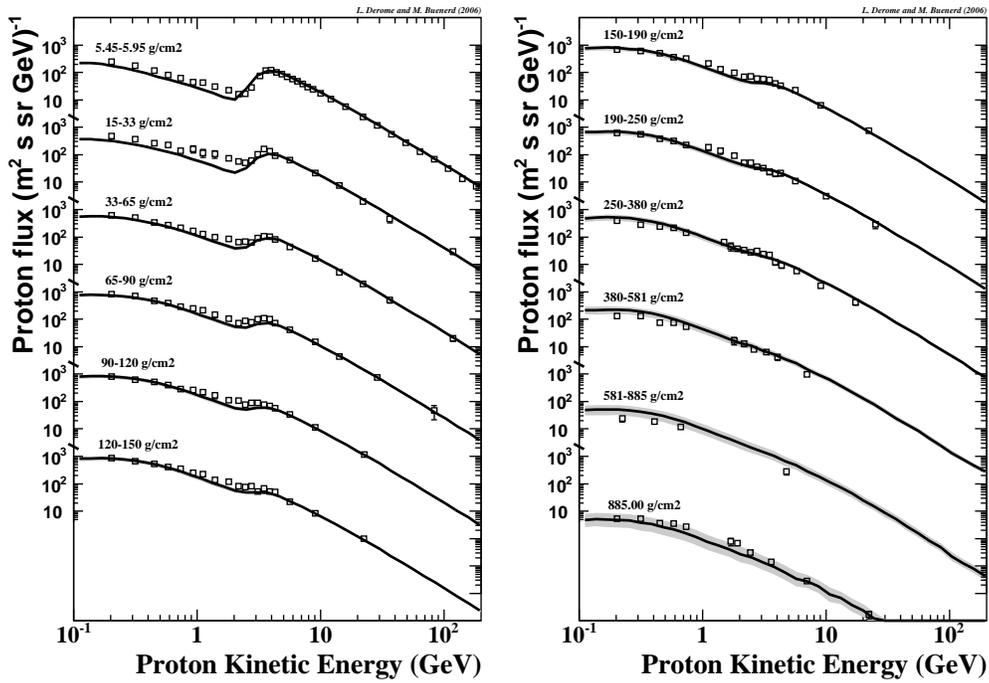}
\caption{\label{CapProton} Proton flux measured by the Caprice experiment (open squares) for different 
altitudes in the atmosphere \cite{CAP98PROTON} compqred with the flux
estimated from the simulation (solid line). The light gray band shows
the 95~\% confidence interval from the simulation. This estimation of
the confidence interval includes only errors from the production cross
sections.}
\end{center}
\end{figure*}  

\section{Results}
\label{sec:results}
\subsection{Proton flux in the atmosphere}

The ability of the simulation to account for the proton production in the atmosphere can be probed by 
comparing the simulation results with the recent $p$ flux measurements performed by the CAPRICE experiment 
between sea level and high altitude ($\sim$ 40 km) \cite{CAP98PROTON}, and by the BESS experiment at mountain 
altitude \cite{BESS99}. 

Figure \ref{bessProton} shows the flux measured by BESS \cite{BESS99} compared to the calculated flux. The 
Bess measurements were performed at Mt Norikura, Japan (geographical location 36N,137.5E) at  2770~m of altitude 
above sea level (corresponding to an atmospheric depth of 742 g/cm2 ) in 1999.

The virtual detector area for this computation was chosen as the geographical region defined by 
$0.4 < |\lambda_{CGM}| < 0.6$~rad where $\lambda_{CGM}$ is the CGM latitude \cite{CGM}, thus consisting of 
two bands (one in each hemisphere) centered on plus or minus the experiment CGM latitude (0.509 rad).

As in the measurements, the geometrical efficiency is set to one for zenithal angular range $\theta_z$ with 
$\cos \theta_z \ge 0.95$.

Figure \ref{bessProton} shows the confidence interval on the accuracy of the simulated flux (with a 95~\% 
level). The estimation of the confidence interval includes only the error associated to the cross section  
and were computed using the confidence interval associated to the cross section parametrization (see section 
\ref{sec:param}). The large width of the confidence interval in this estimation is due to the large number 
 of collisions between the primary particle and the detected proton at
 low altitude. Any uncertainty in the proton production cross section
 is then amplified by a large factor (approximatively 8 on the
 average, see \cite{BA03}).

Figure \ref{CapProton} shows the flux measured by CAPRICE for different altitudes in the atmosphere during 
the 1998 flight from Fort Summer, N. M., (34.28 N, 105.14 W, CGM latitude 0.76 rad) \cite{CAP98PROTON} and 
the corresponding simulation results.

The virtual detector area for this computation was chosen as the region defined by $0.74<|\lambda_{CGM}|<0.78$~rad. 
The geometrical efficiency used here was taken from \cite{CAP98PROTON} to reproduce the detector acceptance.
%if(coszenith<0.84) return 0;
%  if(coszenith<0.88)  return .28;
%  if(coszenith<0.92)  return .45;
%  if(coszenith<0.96)  return .7;
At high altitude, the effect of the geomagnetic cutoff is clearly observed in all spectra. At this CGM 
latitude the mean geomagnetic cutoff is 6 GeV for protons. Above the cutoff the flux is dominated by 
the primary cosmic component whereas below the geomagnetic cutoff the flux consists only of secondary protons
produced by cosmic rays above the cutoff. The primary flux is increasingly absorbed in the atmosphere with the 
decreasing altitude, and the secondary component becomes dominant over the entire energy range.

The confidence interval of the calculated flux is seen to increase with the decreasing altitude, i.e., with the 
mean number of interactions as it could be expected \cite{BA03}. It can be observed that at low altitudes, the 
uncertainty of the simulation results (which includes only the production cross section uncertainties) is much 
larger than the uncertainty from the flux measurements. Therefore, the precise proton flux data at low altitudes 
can be used to constrain the parametrization of the secondary particles production more tightly and more 
accurately than the available nuclear data.
\begin{figure*}
\includegraphics*[width=0.7\textwidth]{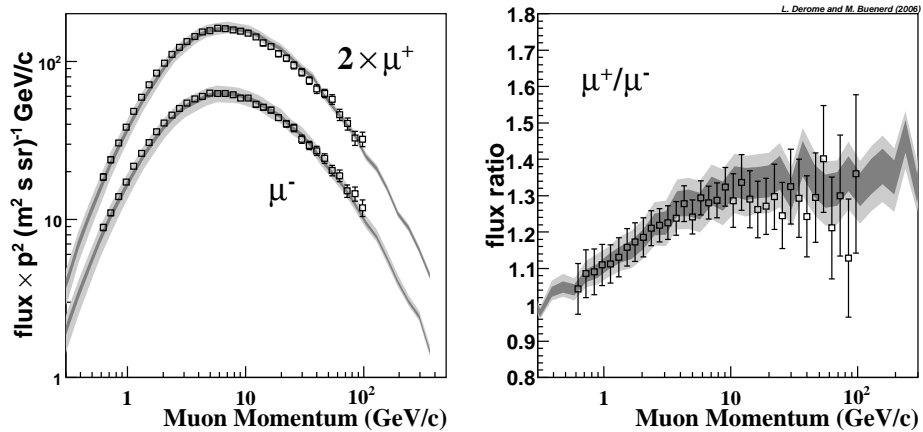}
\caption{\label{fig:BessMountain} Open squares : Muon flux and muon flux ratio measured at Mt Norikura, Japan (location 
36 $^o$N,137.5 $^o$E, altitude 2770 m above see level corresponding to a atmospheric depth of 742 g/cm2 ). Data 
from \cite{BessMuonMountain}. The light gray band shows the 95~\% confidence interval from the simulation. The
estimation of the confidence interval includes only errors from the production cross sections. The dark gray band
corresponds to the  95~\% confidence interval obtained by fitting the cross section on atmospheric data 
(see text, section \ref{sec:atmfit}).}
\end{figure*}
\begin{figure*}
\includegraphics*[width=0.7\textwidth]{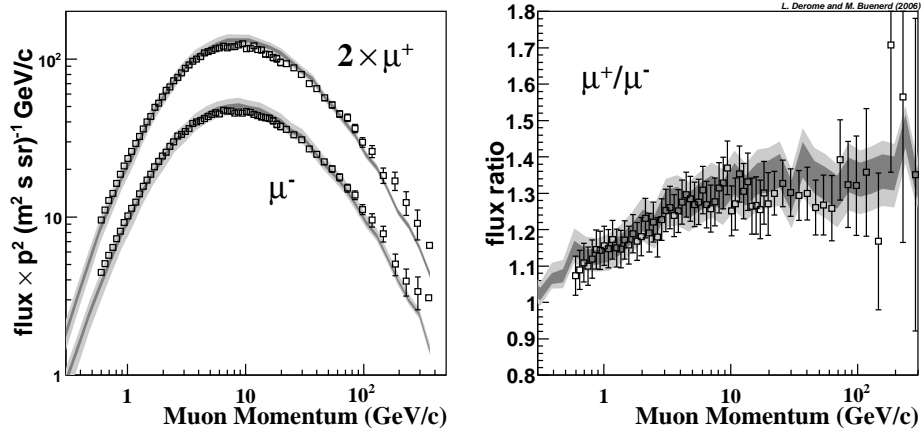}
\caption{\label{fig:BessTsukuba} Open squares : Muon Flux and muon flux ratio in Tsukuba, Japan (location 
36.2 $^o$N,140.1 $^o$E, altitude 30~m above sea level). The data are from \cite{BessMuonTsukuba}.  The light and 
dark gray bands are defined as in figure~\ref{fig:BessMountain}.}
\end{figure*}
\begin{figure*}
\includegraphics*[width=0.4\textwidth]{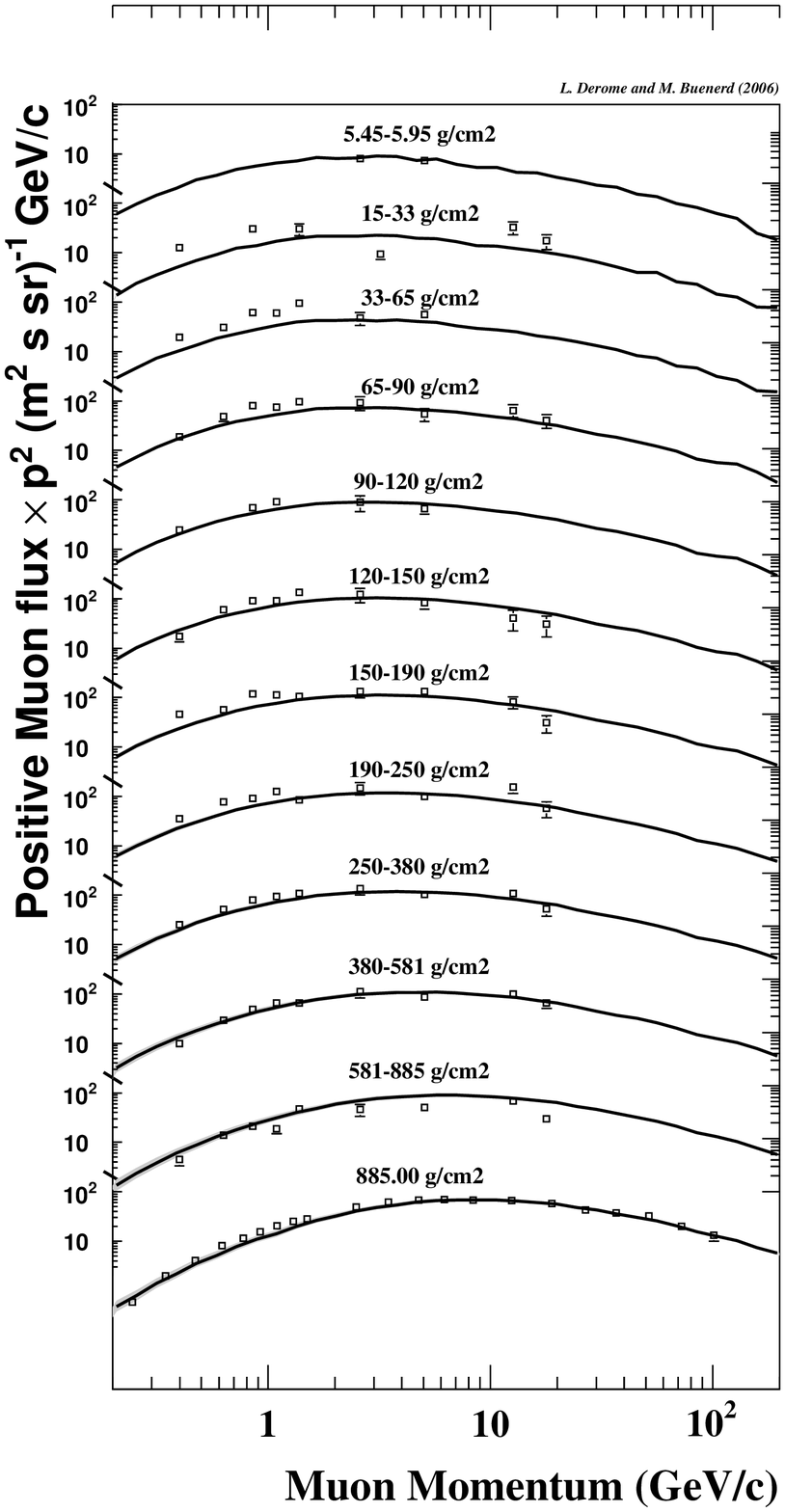}
\hspace{1.5cm}
\includegraphics*[width=0.4\textwidth]{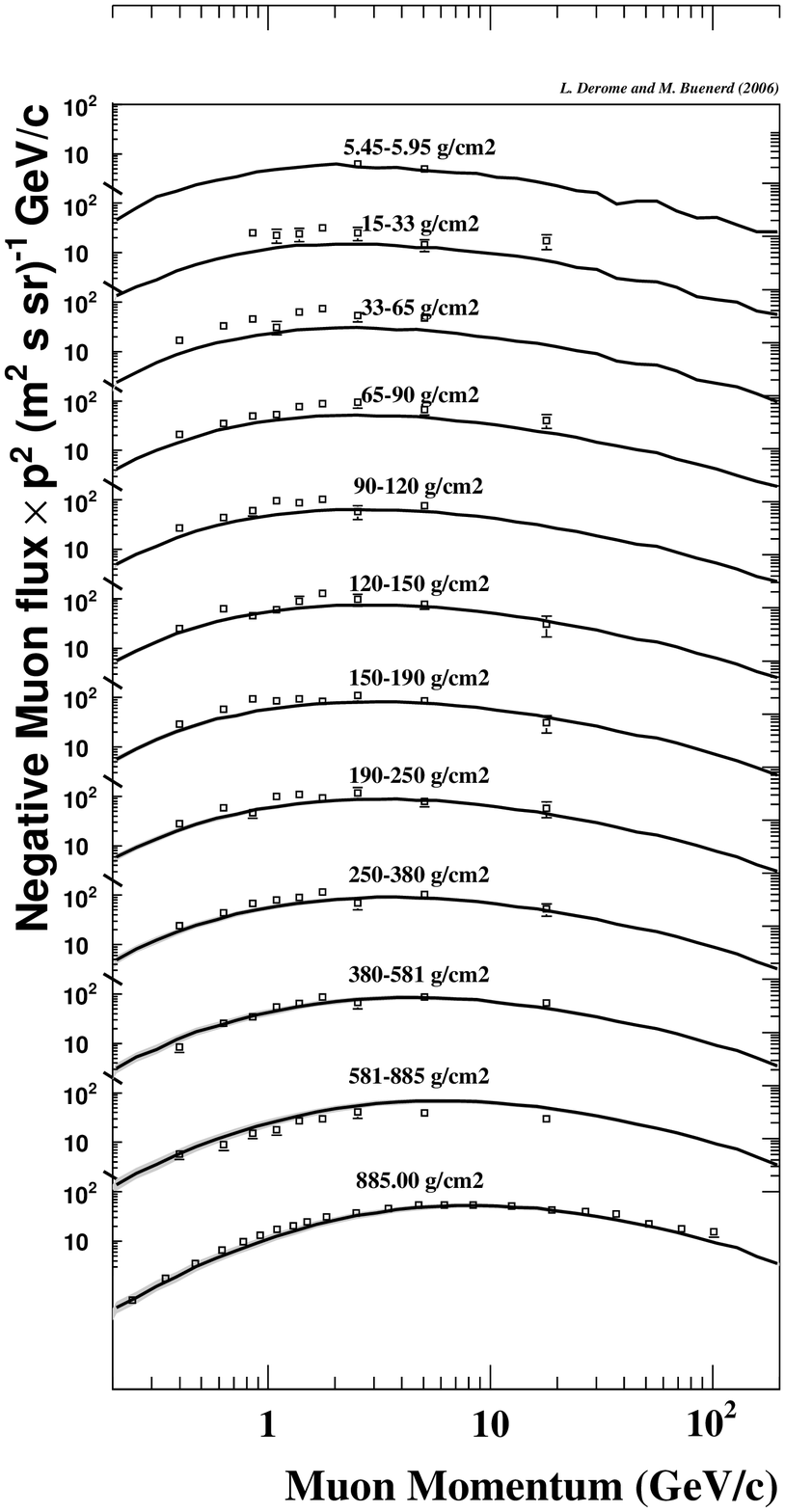}
\caption{\label{fig:Cap98Muon} Positive and negative muon flux measured by the Caprice experiment for different 
altitudes in the atmosphere. The solid line shows the calculated flux and the light gray area corresponds to the 
95\% confidence interval.}
\end{figure*}
\begin{figure*} %[hb]
\begin{center}
\includegraphics[width=0.75\textwidth]{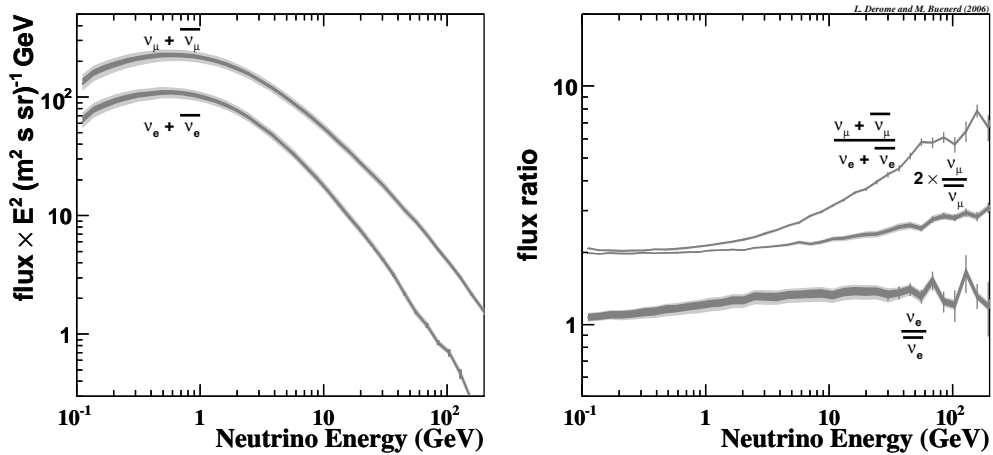}
\caption{\label {fig:flux} Neutrino flux (left) and Neutrino flux ratio (right) at the SK location average over 
$4\pi$. The light gray band corresponds to the 95~\% confidence interval from the simulation. The estimation of the 
confidence interval includes only errors from the production cross sections. The dark gray band represents the 
95~\% confidence interval obtained by fitting the cross section on atmospheric data  (see text, 
section~\ref{sec:atmfit}).}
\end{center}
\end{figure*}
\begin{figure*} %[htb]
\begin{center}
\includegraphics[width=0.4\textwidth]{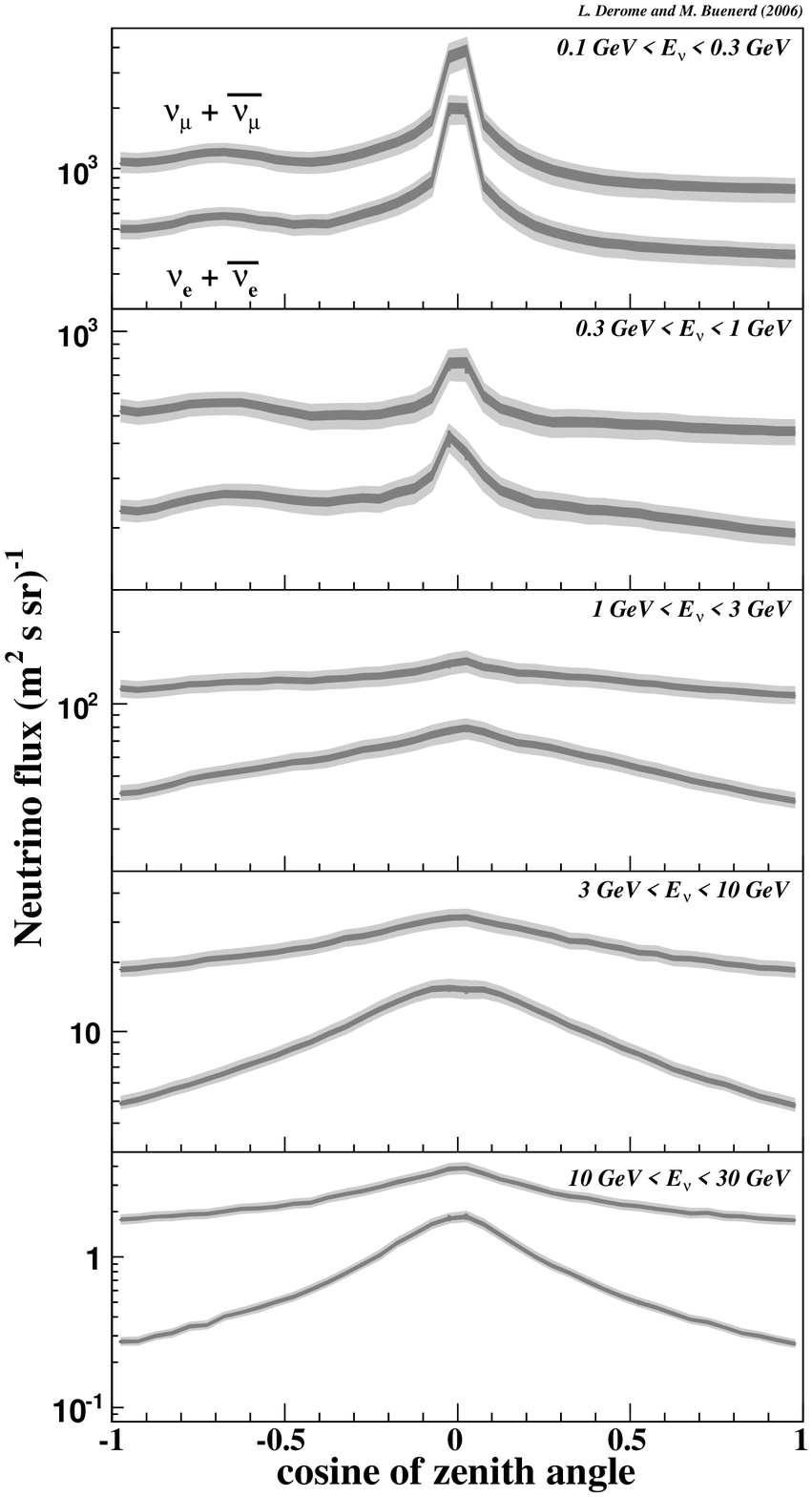}\hspace{1.5cm}\includegraphics[width=0.4\textwidth]{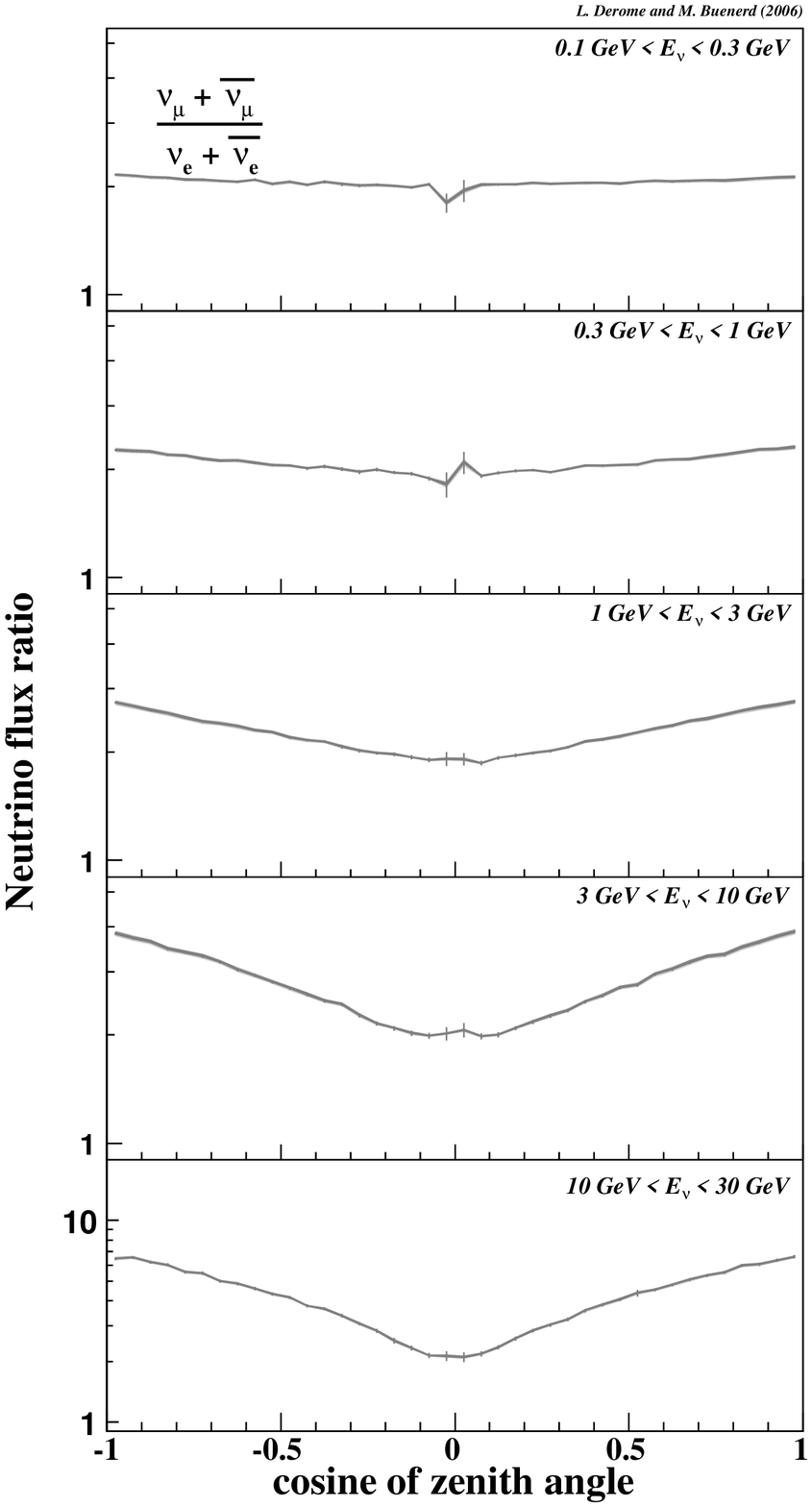}
\caption{\label{zenith} Zenith angle distributions and flavor ratio $(\nu_\mu+{\bar\nu_\mu})/(\nu_e+{\bar \nu_e})$ 
of the calculated neutrino flux  around the Super-Kamiokande detector for five energy bin from 0.1 to 30~Gev. The 
light and dark gray bands are defined as in Fig.~\ref{fig:flux} 
%shows the 95~\% confidence interval obtained from the simulation. The evaluation of the confidence 
%interval includes only errors from the production cross sections. The dark grey band shows the 95~\% confidence 
%interval obtained by fitting the cross sections on the available atmospheric data 
(see text, section \ref{sec:atmfit}).}
\end{center}
\end{figure*}
\begin{figure*} %[htb]
\begin{center}
\includegraphics[width=0.4\textwidth]{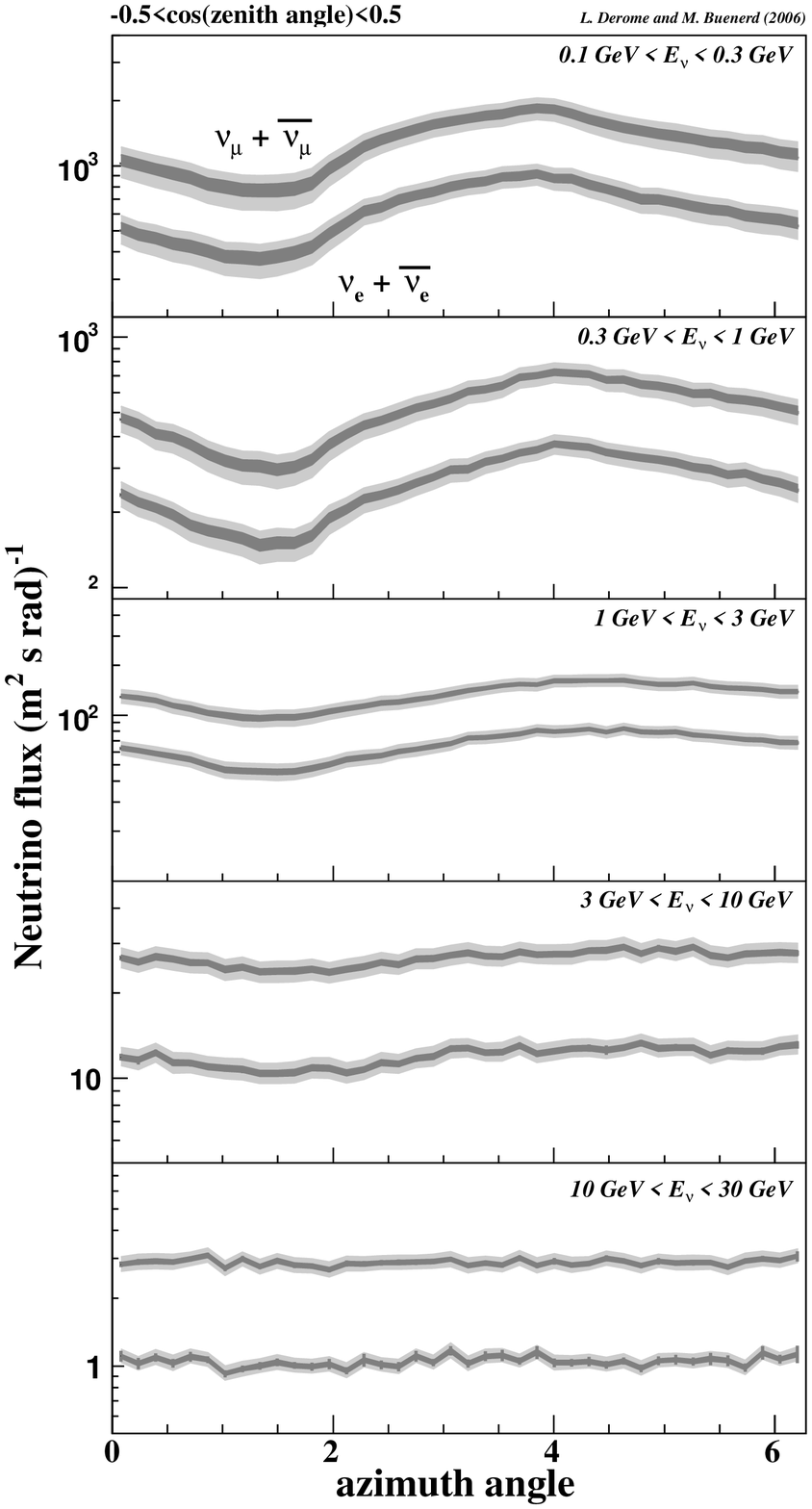}\hspace{1.5cm}\includegraphics[width=0.4\textwidth]{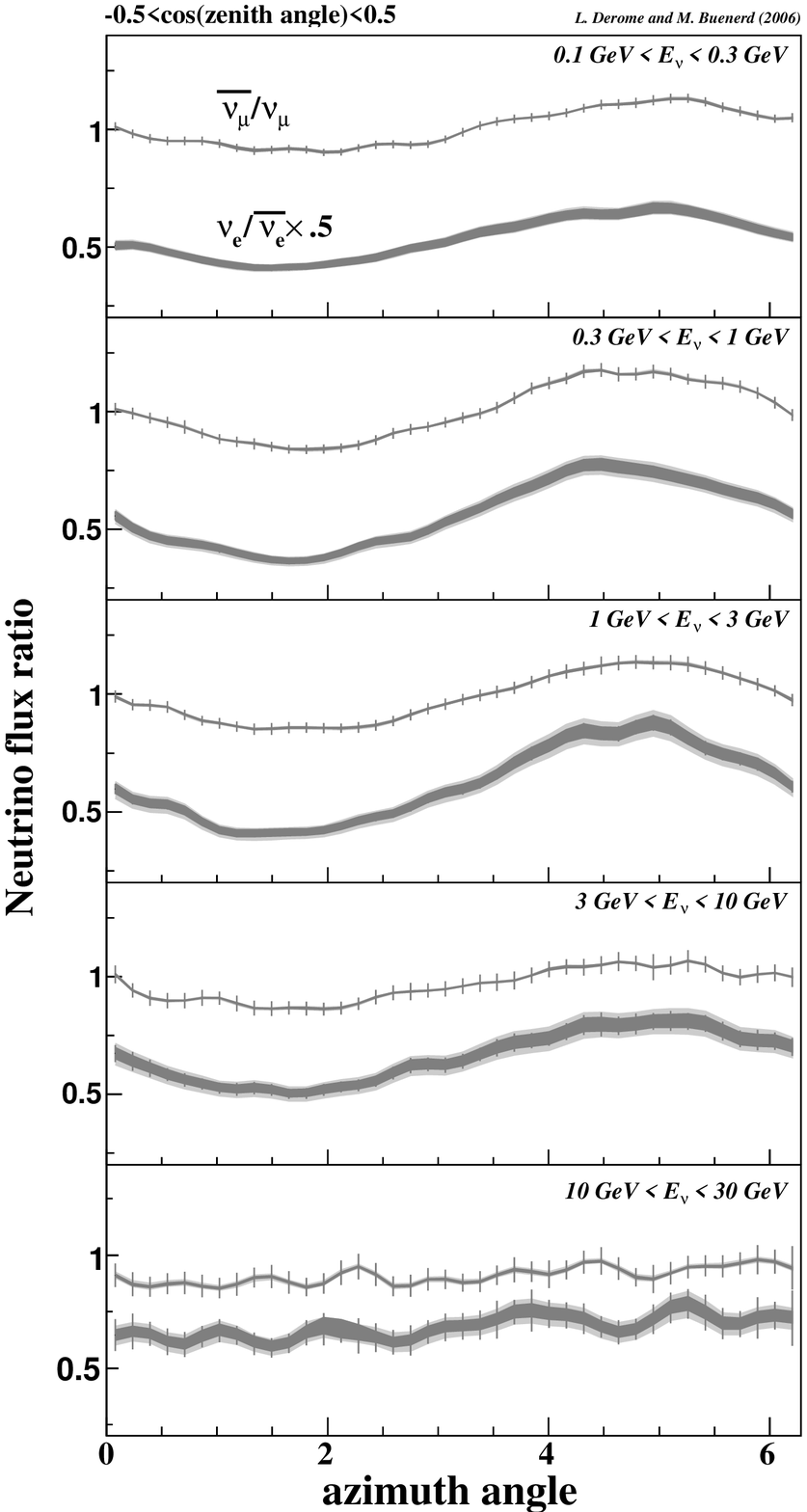}
\caption{\label{azimuth}  Azimuth angle distribution of the neutrino flux (left) and flavor ratios, $\nu_e/{\bar \nu_e}$ 
and ${\bar \nu_\mu}/{\nu_\mu}$  (right) at the SK detector location for five energy bins from 0.1 to 30~Gev. The light and dark 
gray bands are defined as in Fig.~\ref{fig:flux}
%shows the 95~\% confidence interval obtained from the simulation. The estimation of the confidence interval includes only
%errors from the production cross sections. The dark grey band shows the 95~\% confidence interval obtained by fitting the 
%cross section on atmospheric data  
(see text, section \ref{sec:atmfit}).}
\end{center}
\end{figure*}

\subsection{Muon flux in the atmosphere}

Atmospheric muons are produced in the same decay chain as neutrinos. Their spectra are thus an essential ground
to probe the reliability of the neutrino flux calculated in the same framework. Several experiments, CAPRICE, HEAT, 
BESS, have made precise flux measurements at various altitudes in the atmosphere 
\cite{BessMuon2,BessMuonMountain,BessMuonTsukuba,Cap98Muon,CapMuon,HeatMuon} which can be used for this purpose.

Figures \ref{fig:BessMountain} and \ref{fig:BessTsukuba} show the calculated positive and negative muon flux and 
the corresponding flux ratios compared with the data measured by the BESS experiment at mountain altitude (Mt
Norikura, location 36N,137.5E, altitude 2770~m) \cite{BessMuonMountain} and at sea level (Tsukuba, location 36.2
$^o$N,140.1 $^o$E, altitude 30~m) \cite{BessMuonTsukuba}.

For these two calculations, the virtual detectors were chosen as the region defined by $0.4<|\lambda_{CGM}|<0.6$~rad 
at the experiment altitude, and with the geometrical efficiency set to one for zenithal angles 
$\theta_z$ with $\cos \theta_z \ge 0.98$.

The 95 \% confidence interval is also shown on the figure. It includes the uncertainty originating from the 
parametrization of the secondary protons, neutrons and pions production cross sections. Note that for the muon 
charge ratio the uncertainty results only from the uncertainty on the pion production cross section.

%*********** old fig 6 place**********

Figure \ref{fig:Cap98Muon} shows the calculated positive and negative muon flux compared with the data measured by 
the CAPRICE experiment during the 1998 flight \cite{Cap98Muon} at Fort Summer. A good agreement is found between 
data and the calculation for the positive and negative fluxes at different altitude going from the top of atmosphere 
to the ground level.

\subsection{Use of atmospheric flux measurements to constrain hadron production cross sections}
\label{sec:atmfit}
It is clearly seen on the figures \ref{bessProton} to  \ref{fig:Cap98Muon}
that, at least for the low altitude measurements, the statistical precision achieved by the atmospheric 
experiments is significantly better than the precision obtained from the simulation, taking into account the 
uncertainty on secondary particles production cross section. This results was obtained although all the relevant 
available nuclear data have been collected to constrain the parametrized cross sections used in these calculations. 
Therefore, the atmospheric flux measurements are in this sense more accurate that the nuclear data.

In other works on neutrino flux calculations, the calculated atmospheric flux of charged particles, used to check the 
reliability of the calculations, were all based on existing hadronic Monte Carlo generator  \cite{honda,battistoni,barr}, 
whereas in the present approach, in account of the good accuracy of the available measurements, the atmospheric data 
flux can be used to constrain the hadronic cross sections to ultimately improve the final accuracy of the calculated 
neutrino flux.

In this work, the proton and muon measurements at the Mt Norikura altitude and the muon measurements in Tsukuba, from 
the Bess experiment, were used to constrain both the proton and pion production cross sections. The obtained 95~\% 
interval of confidence is shown (dark gray) on figure \ref{bessProton},\ref{fig:BessMountain} and \ref{fig:BessTsukuba}. 
The reduced $\chi^2$ obtained from this fit was 4.7, and the same procedure as described in section \ref{sec:param} 
was thus used to estimate the confidence interval. 
It must be noted that these fitted cross sections may absorb systematic errors originating from other sources, 
like the primary CR fluxes, although the uncertainty on the latter are much smaller than on the cross sections. 
However, this does not affect the ultimate improvement of the neutrino flux uncertainty, since the deduced interval of
confidence accounts for both sources of uncertainty. 
\subsection{Neutrino flux at SuperKamiokande Location}
%*********** old figs 7 8 9 place
The calculated energy distributions of the atmospheric neutrino flux around the SuperKamiokande (SK) detector 
have been computed with the same simulation program. A virtual detector center was defined at the SK 
geographical coordinates (36$^o$N,137$^o$E) with a size of 8$^o$ in latitude and 18$^o$ in longitude 
(or $\sim$900$\times$1600 km$^2$). The virtual detector size has been chosen so as not to change the 
calculated flux by more than 1~\%.

%\todo{Specific Variance reduction techniques used to compute the neutrino flux}

Figure \ref{fig:flux} shows the average flux over the full $4\pi$ range of solid angle for 
$\nu_\mu+\bar{\nu}_\mu$ and $\nu_e+\bar{\nu}_e$ (left) and the flux ratio $\frac{\nu_\mu+\bar{\nu}_\mu}
{\nu_e+\bar{\nu}_e}$, $\frac{\nu_\mu}{\bar{\nu}_\mu}$ and $\frac{\nu_e}{\bar{\nu}_e}$ (right). The light gray area 
represents the 95~\% confidence level corresponding to the uncertainty due to the particle production cross sections 
inaccuracies. The uncertainty for the absolute neutrino flux is of the order of 10~\% and, as expected, these error 
contributions are largely reduced for the ratio of flux $\frac{\nu_e}{\bar{\nu}_e}$ (insensitive to proton/neutron 
production cross sections) and for $\frac{\nu_\mu}{\bar{\nu}_\mu}$ (insensitive to proton/neutron production cross 
sections but also to pion production cross sections for low energy neutrinos) and vanished for the 
$\frac{\nu_\mu+\bar{\nu}_\mu}{\nu_e+\bar{\nu}_e}$ flux ratio (insensitive to proton/neutron and pion production 
cross sections). The dark gray band correspond to the 95~\% confidence interval obtained by fitting the cross 
section on atmospheric data discussed in the section \ref{sec:atmfit}. Using this method the uncertainty for the 
absolute flux is reduced to the order of 3~\%.

The zenith angle distribution of the flux and of the flavor ratios are shown on figure~\ref{zenith}.left 
for 5 energy bins between 0.1 and 30~GeV. At low energy the zenith angle distribution displays an
enhancement for directions close to horizontal, in qualitative agreement with \cite{lip1}. This enhancement 
originates from the isotropisation of the neutrino flux resulting from cosmic ray showers. It is
expected to be maximum for a purely isotropic neutrino emission and nil for a perfectly collimated neutrino 
production. The maximum is then naturally obtained for the lowest energy bin (about twice the downward
or upward going flux out of the horizontal plane). It is decreasing with the increasing energy. At low energy 
the zenith angle distribution for downward and upward neutrinos, outside the $\cos\theta=0$ peak, 
is approximatively flat, this isotropy naturally originating from the CR isotropy. The small structures 
observed arise from geomagnetic effects on the primary flux.
At high energies the distributions become sensitive to the muon decay probability which depends on the muon 
pathlength in the atmosphere and on its energy. As expected the $|\cos \theta_z |$ dependence is becoming 
steeper with the increasing energy.  The effect is enhanced for electronic neutrino which originate exclusively 
from muon decay. This is shown on figure~\ref{zenith}.right where the flavor ratio 
$(\nu_\mu+{\bar\nu_\mu})/(\nu_e+{\bar \nu_e})$ is seen to depend on the zenith angle, especially for high energy 
neutrinos, for the same reason as given previously.

On these figures, the light gray area represents the 95\% confidence level corresponding to the particle 
production uncertainty. Here again we see that the uncertainty is reduced at high energy, this is mainly due 
to the decrease of the mean number of interactions between the primary cosmic ray impinging the atmosphere 
and the final neutrino detection. As in figure~\ref{fig:flux}, for the flavor ratio 
$(\nu_\mu+{\bar\nu_\mu})/(\nu_e+{\bar \nu_e})$ the error contributions from particle production uncertainty vanished.
The dark gray band represents the 95~\% confidence interval obtained by fitting the cross section on atmospheric data as described in section~\ref{sec:atmfit}.

The azimuth angle distribution of the flux and of the flavor ratios are shown on figure~\ref{azimuth} for 5 
energy bins between 0.1 and 30~GeV. On this figure northward and westward going particles correspond to 
azimuth angles 0, $\frac{\pi}{2}$ respectively.
The non-flatness of the low energy bin originates from the geomagnetic cutoff on primary cosmic rays 
\cite{lip2}. The east-west (EW) effect of the primary flux (more eastward going than westward going 
cosmic rays) results in a EW asymmetry in the produced secondary particles and then in the neutrino
flux at low energy. With the increasing energy, neutrinos are increasingly produced in atmospheric showers 
initiated by higher energy cosmic rays which are less and less sensitive to the geomagnetic cutoff (the 
cosmic ray flux becomes isotropic at a rigidity of 60 GV). This results in the simulated azimuthal 
distribution becoming flat for the last energy bin in figure \ref{azimuth}.left (neutrino energy between 
10~Gev and 30~Gev).

For the azimuth angle distributions of the flavor ratios (figure \ref{azimuth}.right), it can be
noted that the $\nu_e/{\bar \nu_e}$ ratio is strongly EW asymmetric
while the opposite is observed for the $\nu_\mu/{\bar\nu_\mu}$
ratio. The $(\nu_\mu+\bar{\nu_\mu})/(\nu_e+\bar{\nu_e})$ ratio (not
shown) is found to be almost structureless at all latitudes. This
additional EW asymmetry featured by the $\nu_e/{\bar \nu_e}$ and
$\nu_\mu/{\bar \nu_\mu}$ ratio can be explained by the muon bending in
the geomagnetic field \cite{lip2}. In the production chain
$p\rightarrow\pi^{+} \rightarrow\mu^{+}\nu_\mu\rightarrow\nu_e{\bar\nu_\mu}$, the muon 
will propagate with the same bending as the proton one. An enhancement of
EW effect originating from the geomagnetic cutoff is then expected for the neutrino $\nu_e$ and 
${\bar\nu_\mu}$ produced in the $\mu^{+}$ decay. In the $p \rightarrow \pi^{-} \rightarrow\mu^{-}{\bar 
\nu_\mu}\rightarrow\nu_\mu{\bar\nu_e}$ production chain,  the muon 
will propagate with the opposite bending and a reduction of the EW
asymmetry is expected for the $\mu^{-}$ decay products
$\nu_\mu$ and ${\bar\nu_e}$. This explains the EW features seen in the
$\nu_e/{\bar \nu_e}$ and the $\nu_\mu/{\bar\nu_\mu}$ ratio. In
addition, because muonic neutrinos are also produced in the pion
decay, the EW asymmetry is found to be higher for $\nu_e/{\bar\nu_e}$
than for ${\bar\nu_\mu}/\nu_\mu$. These results confirm the qualitative prediction
from \cite{lip2}~: 
$$ A_{\bar \nu_e} < A_{\nu_\mu} < A_{\bar \nu_\mu} < A_{ \nu_e}$$
where $A$ stands for the neutrino EW asymmetry.

In figure~\ref{azimuth} the light and dark gray area represents
respectively the 95~\% confidence level corresponding to the particle
production cross section uncertainty and the 95~\% confidence interval
obtained by fitting the cross section on atmospheric data as described
in section~\ref{sec:atmfit}. The uncertainty are largely reduced for
the ratio of flux $\frac{\nu_e}{\bar{\nu}_e}$ (insensitive to the
proton/neutron production cross sections) and for
$\frac{\bar{\nu}_\mu}{\nu_\mu}$ (insensitive to proton/neutron
production cross sections but also to pion production cross sections
for low energy).

\section{Summary and conclusion}
\label{sec:conc}
A full three-dimensional simulation has been used to compute the
secondary particles production in the atmosphere. In these
calculations the statistics of the simulated sample has been
significantly increase with respect to our previous work, resulting in
a negligible contribution of the statistical uncertainty. The dominant
error then appeared to come from the uncertainty on the particle
production cross sections. In this work this systematic uncertainty
has been studied quantitatively.

The simulation has been used to reproduce the atmospheric measurements
of the proton and muon flux, in order to test the reliability of the
calculations. These atmospheric measurements, in account of their good
accuracy, were also used to constrain the secondary particle
production cross sections, which experimental values from direct
measurements on accelerators were of lesser accuracy.

The results for the absolute neutrino fluxes and their zenithal and
azimuthal angular distributions together with the flavor ratio of
these flux at the Super-Kamiokande location, have been presented. The
high statistics accumulated allowed to investigate the detailed
structure of these distribution and to discuss them. The uncertainty
of the calculations has been estimated. It has been shown first, that
this uncertainty is at the level of 10~\% for the absolute flux, and
second that it can be reduced to the level of 3~\% by using the
atmospheric data on the muon and proton flux to constrain the
production cross sections. These results provide an accurate basis for
the comprarison to future experimental data.

\begin{acknowledgments}
The author is very grateful to Michel Bu\'enerd without whom this work would not have been initiates and for his precious help in writing the manuscript.
\end{acknowledgments}

\end{document}